\documentclass[entropy,article,accept,moreauthors,pdftex,10pt,a4paper]{mdpi}
\usepackage{amssymb}
\usepackage{amsmath}
\usepackage[figuresright]{rotating}
\usepackage{todonotes}

\usepackage{algorithmic}							
\usepackage[lined,boxed,ruled,commentsnumbered]{algorithm2e}						
\usepackage{hyperref}          

\firstpage{1}
\articlenumber{x}
\doinum{10.3390/------}
\pubvolume{xx}
\pubyear{2016}
\copyrightyear{2016}
\externaleditor{Academic Editor: Kevin H. Knuth}
\history{Received: 1 August 2016; Accepted: 14 October 2016; Published: date}
\pdfoutput=1

\Title{Point Information Gain and Multidimensional Data~Analysis}

\Author{\mbox{Renata Rycht\'{a}rikov\'{a} $^{1,}$*, Jan Korbel $^2$, Petr Mach\'{a}\v{c}ek $^1$, Petr C\'{i}sa\v{r} $^1$, Jan Urban $^1$ and Dalibor \v{S}tys $^1$}}
\AuthorNames{Renata Rycht\'{a}rikov\'{a}, Jan Korbel, Petr Mach\'{a}\v{c}ek, Petr C\'{i}sa\v{r}, Jan Urban and Dalibor \v{S}tys}

\address{%
$^{1}$ \quad Institute of Complex Systems, South Bohemian Research Center of Aquaculture and Biodiversity of Hydrocenoses, Faculty of Fisheries and Protection of Waters (FFWP),
 University of South Bohemia in \v{C}esk\'{e} Bud\v{e}jovice, Z\'{a}mek 136, Nov\'{e} Hrady 373 33, Czech Republic; info@imagecode.eu (P.M.);
 cisar@frov.jcu.cz (P.C.); urbanj@frov.jcu.cz (J.U.); stys@frov.jcu.cz (D.S.)\\
$^{2}$ \quad  Faculty of Nuclear Sciences and Physical Engineering, Czech Technical University in Prague, B\v{r}ehov\'{a} 7, Prague  155 19, Czech Republic; korbeja2@fjfi.cvut.cz}

\corres{Correspondence: rrychtarikova@frov.jcu.cz; Tel.: +420-387-773-833}

\abstract{We generalize the point information gain (PIG) and derived quantities, i.e.,  point information gain entropy (PIE) and point information gain entropy density (PIED), for the case of the R\'{e}nyi entropy and simulate the behavior of PIG for typical distributions. We also use these methods for the analysis of multidimensional datasets.  We demonstrate the main properties of PIE/PIED spectra for the real data with the examples of several images and discuss further possible utilizations in other fields of data processing.}


\begin{document}

\section{Introduction}
Measurement of relative information between two probability distributions is one of the most important goals of information theory. Among many other concepts, there are two that are widely used. By far, the most widespread concept is called the relative Shannon entropy or the Kullback--Leibler divergence. In this work, we use an alternative approach based on a simple concept of entropy difference instead. By generalization of both concepts from Shannon's approach to R\'{e}nyi's approach, we obtain the whole class of information variables that enable aiming for different parts of probability distributions and interpret it as an investigation of different parts of multifractal systems.

Despite the mathematical precision of the concept of the Shannon/R\'{e}nyi divergence, we use another concept, the (R\'{e}nyi) entropy difference, for introduction of a value which locally determines an information contribution of a given element in a discrete set. Even though there is no substantial restriction on the usage of a standard divergence for calculation of the information difference upon elimination of one element from a set, for practical reasons, we used the simple concept of entropy difference between sets with and without a given element. The resulted value has been called a point information gain $\Gamma_{\alpha}^{(i)}$~\cite{Ur08,Ur09}. The goal of this article is to examine and demonstrate some properties of this variable and derive another quantities, namely a point information gain entropy $H_\alpha$ and a point information gain entropy density $\Xi_\alpha$. We also introduce the relation of all these variables to global and local information in multidimensional data analysis.
\section{Mathematical Description and Properties of Point Information Gain}

\subsection{Point Information Gain and Its Relation to Other Information Entropies}\label{subsec:PIG}

An important problem in the information theory is to estimate the amount of information gained or lost by refining, and approximate the probability distribution $P$ by the distribution $Q$. The most popular measure used in the theory is the Kullback--Leibler (KL) divergence, defined as
\begin{equation}
D_{KL}(P||Q) = \sum_j p_j \ln \frac{p_j}{q_j} = E_p \left[\ln P\right] - E_p \left[\ln Q\right] = S_P(Q) - S(P),
\end{equation}
where $S_P(Q)$ is so-called cross-entropy~\cite{cross} and $S(P)$ is the Shannon entropy of distribution $P$. If $P$ is similar to $Q$, this measure can be approximated by entropy difference
\begin{equation}
\Delta S(P,Q) = S(Q) - S(P).
\end{equation}
Indeed, this measure does not obey as many theoretic-measure axioms as the KL-divergence. For~ instance, for $P \neq Q$, we can still obtain $\Delta S(P,Q) = 0$. Nevertheless, if $P \approx Q$ and $P \neq Q$, this value can be still a suitable quantity revealing some important information aspects of a system. The situation, when the distributions are approximative histograms of some underlying distributions $P$ for $n$ and ($n+1$) measurements, respectively, is particularly interesting. In this case, the entropy~difference
\begin{equation}
\Delta S(P_n,P_{n+1}) = S(P_{n+1}) - S(P_{n})
\end{equation}
can be interpreted as an information gained by the $(n+1)$-th measurement. Naturally, $P_n \rightarrow P$.
When dealing with real complex systems, it is sometimes advantageous to introduce new information variables and entropies that capture the complexity of the system better, e.g., Hellinger's distance, Jeffrey's distance or J-divergence. There are also some specific information measures that have special interpretations and are widely used in various applications~\cite{baez2011, marcolli2015}. Two of the most important quantities are the Tsallis--Havrda--Charv\'{a}t (THC) entropy~\cite{Ts88}, which is the entropy of non-extensive systems, and the R\'{e}nyi entropy, the entropy of multifractal systems~\cite{Ji04a, Jizba14}. The latter is tightly connected to the theory of multifractal systems and generalized dimensions~\cite{He85}. It is defined as
\begin{equation}
\mathcal{H}_\alpha(P) = \frac{1}{1-\alpha} \ln \sum_j p_j^{\alpha}, \quad \alpha \geq 0,
 \end{equation}
where $\alpha$ is the R\'{e}nyi coefficient and $p_j$ is the probability of occurrence of a phenomenon $j$ in the discrete distribution. Limit $\alpha \rightarrow 1$ recovers the Shannon entropy.
Similar to the Shannon entropy, the R\'{e}nyi entropy also has an operational meaning. Actually, it can be interpreted as the average information cost, when the cost of an elementary piece of information is an exponential function of its~length~\cite{Ca65}. Thus, changing the parameter $\alpha$ changes the cost of the information and therefore accentuates some parts of the probability distributions while suppressing the others. Thus, by taking into account the whole class of R\'{e}nyi entropies, we get a new generalized class of information~quantities.

The point information gain $\Gamma_\alpha^{(i)}$ 
of the $i$-th point was developed as a practical tool for assessment of the information contribution of an element to a given discrete distribution \cite{St12}. Similar to the Shannon entropy difference, it is defined as a difference of two R\'{e}nyi entropies---with and without the examined element of a discrete phenomenon. Let us consider a discrete distribution of $k$ distinct possible outcomes (e.g., different colors of pixels). Let us have a discrete distribution
\begin{equation}
P = \{p_j\}_{j=1}^k = \left\{\frac{n_1}{n},\dots,\frac{n_i}{n},\dots,\frac{n_k}{n}\right\},
\end{equation}
where $n$ denotes the total number of elements in the discrete distribution and $n_i$ the number of elements of $i$-th phenomenon, $i \in \{1,2, ..., k-1, k\}$, respectively. Let us denote $n_j^{(i)} = n_j$ for $j \neq i$ and $n_i^{(i)} = n_i -1$. Then, the distribution with the omitted $i$-th phenomenon can be written as
\begin{equation}P^{(i)} = \left\{p_{j}^{(i)}\right\}_{j=1}^k = \left\{\frac{n_1^{(i)}}{n-1},\dots,\frac{n_i^{(i)}}{n-1},\dots,\frac{n_k^{(i)}}{n-1}\right\}.
\end{equation}
Hence, we may write the point information gain $\Gamma_{\alpha}^{(i)}$ as
\begin{eqnarray}
\label{eg:1}
\Gamma_{\alpha}^{(i)}& =& \mathcal{H}_{\alpha}\left(P^{(i)}\right) - \mathcal{H}_\alpha(P) = \frac{1}{1-\alpha}\ln{\left(\sum_{j=1}^k{\left(p_{j}^{(i)}\right)^\alpha}\right)} -  \frac{1}{1-\alpha} \ln{ \left(\sum_{j=1}^k{(p_j)^\alpha}\right)} \nonumber\\&=& \frac{1}{1-\alpha} \ln \left(\frac {\sum_{j=1}^k{\left(p_{j}^{(i)}\right)^\alpha}} {\sum_{j=1}^k{p_j^\alpha}}\right),
\end{eqnarray}
where $k$ is the total number of the phenomena in the discrete distribution. In the rest of the text, we use the natural logarithm to simplify calculations. However, all computations have been performed with the usage of binary logarithm which, for the R\'{e}nyi entropy and its derivatives, yields values in bits. 
In contrast to the commonly used R\'{e}nyi divergence~\cite{Re61,Ku57,Cs75,Ha06,Er07,Er10,Ji12}, we use $\Gamma_{\alpha}^{(i)}$ for its relative simplicity and practical interpretation. Unlike the KL divergence, the R\'{e}nyi divergence cannot be interpreted as a difference of cross-entropy and entropy of the underling distribution and computation becomes intractable. As discussed above, for similar distributions, it still preserves its information values.

After the substitution for the probabilities, one gets that
\begin{equation}
\Gamma_{\alpha}^{(i)} = \frac{1}{1-\alpha}\ln{\frac {\sum_{j=1}^k{\frac{\left(n_{j}^{(i)}\right)^\alpha}{(n-1)^\alpha}}} {\sum_{j=1}^k{ \frac{n_j^\alpha}{n^\alpha}}}} = \mathcal{C}_\alpha(n)  + \frac{1}{1-\alpha}\ln{\frac{\sum_{j=1}^k{\left(n_{j}^{(i)}\right)^\alpha}} {\sum_{j=1}^k{n_j^\alpha}}},
\label{eg:3}
\end{equation}
where $\mathcal{C}_\alpha(n) = \ln\left(\frac{n}{n-1}\right)^\frac{\alpha}{1-\alpha}$ depends only on $n$. For $n \rightarrow \infty$ and $\Gamma_{\alpha}^{(i)} \rightarrow 0$,   the whole entropy remains finite (contrary to unconditional entropy, which has to be renormalized for continuous case (for details, see Reference~\cite{Ji04a})). 
Therefore, we examine only the second term. When the argument of the logarithm is close to~1, i.e.,
\begin{equation}
\sum_{j=1}^k \left(n_{j}^{(i)}\right)^\alpha  \approx \sum_{j=1}^k n_j^\alpha,
\end{equation}
which leads to the condition that
\begin{equation}
\left(\frac{n_{i}^{(i)}}{n_i}\right)^\alpha = \left(\frac{n_i - 1}{n_i}\right)^\alpha \approx 1,
\end{equation}
for given $\alpha$, one can then approximate the logarithm by the Taylor expansion of the first order. After~denoting
\begin{equation}
\mathcal{D}_{\alpha}^{(i)} = \frac{\sum_{j=1}^k{\left(n_{j}^{(i)}\right)^\alpha}} {\sum_{j=1}^k{n_j^\alpha}},
\end{equation}
the second term of $\Gamma_{\alpha}^{(i)}$ can be approximated as
\begin{equation}
\frac{1}{1-\alpha}\ln{\mathcal{D}_{\alpha}^{(i)}} = \frac{1}{1-\alpha}\left(\mathcal{D}_{\alpha}^{(i)} -1\right) + \mathcal{O}\left(\left(\mathcal{D}_{\alpha}^{(i)} -1\right)^2\right),
\label{eg:4}
\end{equation}
where we used the big $\mathcal{O}$ asymptotic notation. Let us note that the last term in Equation~\eqref{eg:3} is nothing else than the THC entropy~\cite{Ts88,Ha67}. Naturally, for very similar distributions, these two quantities are practically the same. This is due to the fact that, for large $n$, the omission of the point has no large impact on the whole distribution. Consequently, an actual value of parameter $\alpha$, which leads to rescaling of probabilities, is more important than a particular form of entropy.

We shall continue by utilizing the R\'{e}nyi entropy due to its relation to the generalized dimension of multifractal systems~\cite{Gr83a, Gr83b}. Let us concentrate again to the term $\mathcal{D}_{\alpha}^{(i)}$. We can rewrite it as
\begin{equation}
\mathcal{D}_{\alpha}^{(i)} = \frac{\sum_{j=1}^k{\left(n_{j}^{(i)}\right)^\alpha}} {\sum_{j=1}^k{n_j^\alpha}} = \frac{\sum_{j=1, j\neq i}^{k}{n_j^\alpha}+(n_i-1)^\alpha}{\sum_{j=1}^k{n_j^\alpha}} = 1 - \alpha \frac{n_i^{\alpha-1}}{\sum_{j=1}^k{n_j^\alpha}} + \frac{1}{\sum_{j=1}^k{n_j^\alpha}} \omega \left(n_i^{\alpha-2}\right),
\end{equation}
where we use the small $\omega$ asymptotic notation. Specifically, provided $\alpha=2$, we obtain
\begin{equation}
\Gamma_{2}^{(i)} \approx \mathcal{C}_{2}(n) + \frac{1}{1-2}\left(\frac {\sum_{j=1, j\neq i}^{k}{n_j^2}+(n_i-1)^2} {\sum_{j=1}^k{n_j^2}}-1\right)  \approx \mathcal{C}_{2}(n) + \frac{2 n_i - 1}{\sum_{j=1}^k{n_j^2}},
\label{eg:5}
\end{equation}
which explains why the dependency $n_i$ on $\Gamma_{2}^{(i)}$ is approximately linear. 
In general, 
point information gain is a monotone function of $n_i$, respectively $p_i$, for all possible discrete distributions. Thus, it may be used as a quantity of information gain between two discrete distributions, which in the occurrence of one particular feature, differ.

Let us discuss an interpretation of the point information gain. We can rewrite Equation \eqref{eg:3} as
\begin{eqnarray}
\Gamma_\alpha^{(i)} &=& \ln \left(\frac{n}{n-1}\right)^{\frac{\alpha}{1-\alpha}}  + \ln \left(1 + \frac{(n_i - 1)^\alpha - n_i^\alpha}{\sum_{j=1}^k n_j^\alpha} \right)^\frac{1}{1-\alpha}  \nonumber\\&=& - \ln \left[\left(1-\frac{1}{n}\right)^\alpha\right]^{\frac{1}{1-\alpha}} + \ln \left(1 + n_i^\alpha \frac{\left(1 - \frac{1}{n_i}\right)^\alpha - 1}{\sum_{j=1}^k n_j^\alpha} \right)^\frac{1}{1-\alpha}.
\end{eqnarray}
We are interested in the situation when $\Gamma_\alpha^{(i)} = 0$. After straightforward manipulations, we can get rid of $\ln$ and $\frac{1}{1-\alpha}$ power, so
\begin{equation}
\left(1-\frac{1}{n}\right)^\alpha = 1 + n_i^\alpha \frac{\left(1 - \frac{1}{n_i}\right)^\alpha - 1}{\sum_{j=1}^k n_j^\alpha}.
\end{equation}
If $n \gg 1$ and $n_i \gg 1$, we can approximate both sides with the rule $\left(1+x\right)^\alpha \approx 1+ \alpha x$ for $x$ close to zero which gives
\begin{equation}
1-\frac{\alpha}{n} = 1 - \frac{\alpha n_i^{\alpha-1}}{\sum_{j=1}^k n_j^\alpha}.
\end{equation}
Thus, we end with
\begin{equation}
n_i = \sqrt[\alpha-1]{\frac{\sum_{j=1}^k n_j^\alpha}{n}}.
\end{equation}

This shows that $\Gamma_\alpha^{(i)} = 0$ holds for events with average frequency.
$\Gamma_{\alpha}^{(i)} < 0$ corresponds to rare events, while $\Gamma_{\alpha}^{(i)} > 0$ corresponds to frequent events. Thus, in addition to the definition of the quantity of the contribution of each event to the examined distribution, we also obtain the discrimination between points which contribute to the total information of the given distribution under the statistical assumption represented by a particular $\alpha$. This opens the question on existence of the ``optimal'' distribution for the given $\alpha$.

Then, the possible variants of such optimality arise subsequently: the first one can be defined as a distribution for which exactly half of the value $n_i$ produces $\Gamma_{\alpha}^{(i)} > 0$ and the other half yields $\Gamma_{\alpha}^{(i)} < 0$. The second one requires values $\Gamma_{\alpha}^{(i)}$ to be spaced equally. Existence of such a distribution could be understood as another generalization of the concept of the entropy power~\cite{Costa1985, Jizbaetal2014}, and we refer this question to our future research.

\begin{figure}[H]
\begin{center}
\includegraphics[width=0.98\textwidth]{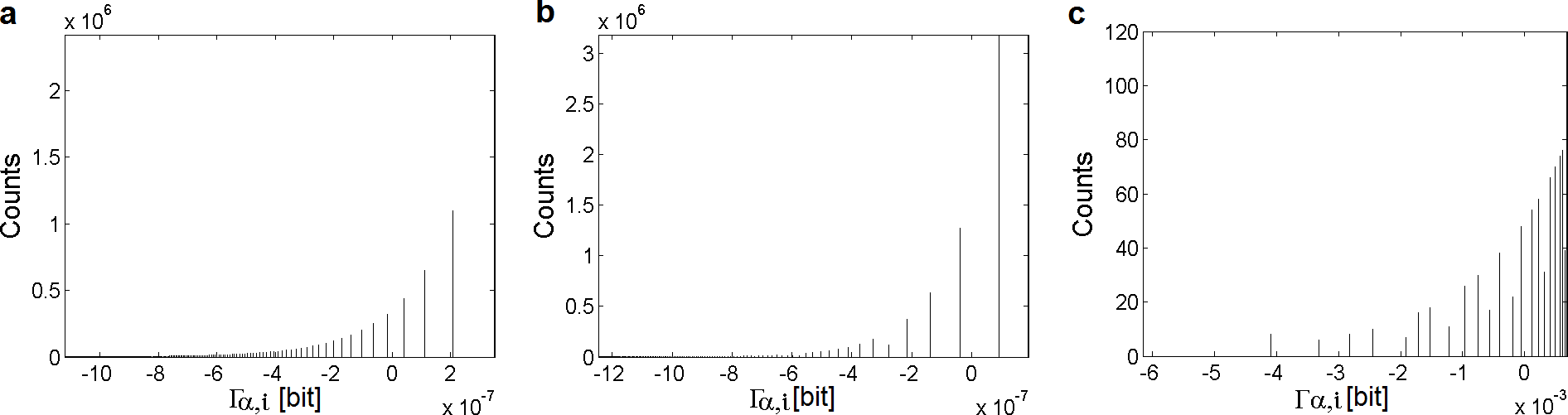}
\end{center}
\caption{$\Gamma_{\alpha, i}$-transformations of the discretized L\'{e}vy (\textbf{a}), Cauchy (\textbf{b}), and Gauss
(\textbf{c}) distribution at $\alpha$ = 0.99. The deviation from the monotone dependency in the Gauss distribution is due to the digital~rounding.}
\label{Fig1}
\end{figure}

With respect to the previous discussion and practical utilization of this notion, we emphasize that for real systems with large $n$, values $\Gamma_{\alpha}^{(i)}$ are relatively small numbers for current numerical precision of common computers. Their further computer averaging and numerical representation lead to significant errors such as underflow and overflow (e.g., Figure~\ref{Fig1}c). At lower values $\alpha$, the values $\Gamma_{\alpha}^{(i)}$ are broadly separated for rare points, while, at higher values $\alpha$, the resolution is higher for more frequent data points. Therefore, spectrum $\Gamma_{\alpha}^{(i)}$ vs. $\alpha$ is more advisable to compute rather than a single $\Gamma_{\alpha}^{(i)}$ value at a chosen $\alpha$.

\subsection{Point Information Gain for Typical Distributions}
\label{subsec:distributions}
In Figure~\ref{Fig1}, we demonstrate $\Gamma_{\alpha}^{(i)}$-transformations of three thoroughly studied distributions---the L\'{e}vy, Cauchy, and Gauss distribution (specified in Section~\ref{image_process}). Mainly,
Figure~\ref{Fig1}c shows averaging of digital levels, which results in multiple appearance of unique points. This phenomenon is reduced with the increasing number of the points in the distribution. Nevertheless, it does not disappear in any real case. Thus, the monotone dependencies of $n_i$, respectively $p_i$, on the $\Gamma_{\alpha}^{(i)}$ are valid only at the approximation to an infinite resolution in levels of values.

\begin{figure}[H]
\begin{center}
\includegraphics[width=\textwidth]{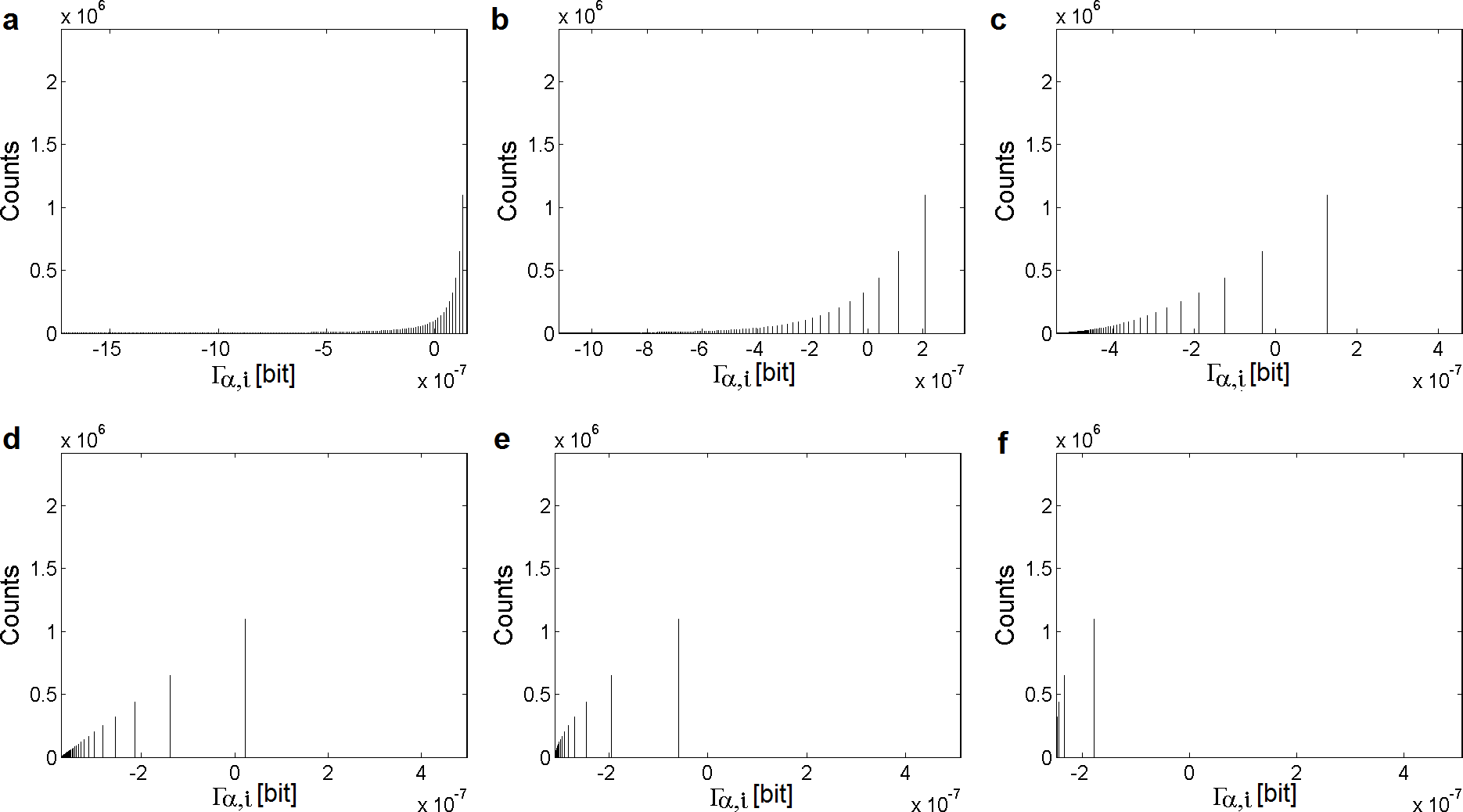}
\end{center}
\caption[]{$\Gamma_{\alpha}^{(i)}$-transformations of the discretized L\'{e}vy distribution at $\alpha$ = $\{0.5, 0.99, 1.5, 2.0, 2.5, 4.0\}$ (from (\textbf{a}) to (\textbf{f})).}
\label{Fig2}
\end{figure}

Figure~\ref{Fig2} shows distribution changes of the values $\Gamma_{\alpha}^{(i)}$ with the increasing $\alpha$-parameter. For each parameter $\alpha$, the elements $\Gamma_{\alpha}^{(i)}$ are enveloped by monotone increasing curves. For instance, as devised in Equation~\eqref{eg:5}, the near linearity of the dependency of the number of elements on the values $\Gamma_{\alpha}^{(i)}$ at $\alpha$ = 2 is seen in Figure~\ref{Fig2}d. The differences between the  distributions are expressed by the distributions of the values $\Gamma_{\alpha}^{(i)}$ along the horizontal axes.

\subsection{Point Information Gain Entropy and Point Information Gain Entropy Density}
\label{subsec:PIE_and_PIED}
In the previous sections, we showed that $\Gamma_{\alpha}^{(i)}$ is different for any $n_i$ and the dependency of these two variables is a monotone increasing function for all $\alpha > 0$.
Here, we propose new variables---a point information gain entropy ($H_\alpha$) and point information gain entropy density ($\Xi_\alpha$) defined by~formulas
\begin{equation}\label{eg:6}
H_\alpha=\sum_{j=1}^k n_j \Gamma_{\alpha}^{(j)}
\end{equation}
and
\begin{equation}\label{eg:7}
\Xi_\alpha=\sum_{j=1}^k \Gamma_{\alpha}^{(j)}.
\end{equation}
They can be understood as a multiple of the average point information gain and---under linear averaging---an average gain of the phenomenon $j$, respectively.

The information content is generally measured by the entropy. The famous Shannon source coding theorem~\cite{Sh48} refers to a specific process of transmission of a discretized signal and introduction of the noise. The R\'{e}nyi entropy is one of the class of one-parametric entropies and offers numerous additional features over the Shannon entropy~\cite{Ji04a, Re61, Ji04b} such as the determination of a generalized dimension of a strange attractor~\cite{Gr83a, Gr83b}. The universality of the generalized dimension for characterization of any distribution, whose regularity may be only coincidental, is still under dispute. However, the values $H_\alpha$ and $\Xi_\alpha$ characterize a given distribution for any $\alpha$. Differences between distributions are expressed in counts along the axes $\Gamma_\alpha$. Therefore, independently of the mechanisms of the generation of the distributions, the values $H_\alpha$/$\Xi_\alpha$ can serve for the comparison of these distributions. It holds for any both parametric and non-parametric distributions.

The next question is whether the $\Xi_\alpha$ has some expected properties. In this aspect, we mention the facts observed upon examination of Equation~\eqref{eg:4}, which enable us to rewrite it as
\begin{eqnarray}
\Xi_\alpha&=&\sum_{j=1}^k \Gamma_{\alpha}^{(j)} = \frac{k}{1-\alpha} \ln \left(\frac{n^\alpha}{(n-1)^\alpha}\right) +  \frac{1}{1-\alpha} \sum_{i=1}^k \ln\left(\frac{\sum_{j=1, j\neq i}^{k}{n_j^\alpha}+(n_i-1)^\alpha} {\sum_{j=1}^k n_j^\alpha}\right) =\nonumber \\ &=&  \mathcal{C}_\alpha(n) \cdot k + \frac{1}{1-\alpha} \ln \left( \prod_{j=1}^k \mathcal{D}_{\alpha}^{(j)} \right),
\label{eg:8}
\end{eqnarray}
where the product in the argument of the logarithm in the second term is a product of functions upper limited by 1 and thus again a function upper limited by $1$. From the previous analysis done for the $\mathcal{D}_{\alpha}^{(i)}$, we may conclude that the point information gain entropy density ($\Xi_\alpha$) inherits properties of R\'{e}nyi entropy, i.e., zooming properties, etc.

Similar to Equation \eqref{eg:8}, the point information gain entropy ($H_\alpha$) can be rewritten as
\begin{eqnarray}
H_\alpha&=&\sum_{j=1}^k n_j \Gamma_{\alpha}^{(j)} = \frac{\sum_{j=1}^k n_j}{1-\alpha} \ln \left(\frac{n^\alpha}{(n-1)^\alpha}\right) + \sum_{i=1}^k n_i \ln \left(\frac{(\sum_{j=1, j\neq i}^{k}{n_j^\alpha}+(n_i-1)^\alpha)} {\sum_{j=1}^k {n_j^\alpha}}\right) = \nonumber \\ &=&  \mathcal{C}_\alpha(n) \cdot \sum_{j=1}^k n_j +  \ln \left( \prod_{j=1}^k \left(\mathcal{D}_{\alpha}^{(j)}\right)^{n_j} \right).
\label{eq:10}
\end{eqnarray}
Again, the argument of the logarithm in the second term is upper limited by $1$. The $H_\alpha$ also has properties inherited from R\'{e}nyi entropy, although their mutual relation is more complicated.

\section{Estimation of Point Information Gain in Multidimensional Datasets}
\subsection{Point Information Gain in the Context of Whole Image}
\label{subsec:PIGmultidimensional}
Point information gain $\Gamma_{\alpha,i}$ introduced in Equation~\eqref{eg:1} was originally applied to the image enhancement~\cite{Ur08,Ur09}. A typical digital image is a matrix of $x \times y  \times n$ values, where $x$ and $y$ are dimensions of the image and $n$ corresponds to the number of color channels (e.g., $n$ is 1 and 3 for a monochrome and RGB image, respectively). In most cases, the intensity values are in the range from 0 to 255 (an 8-bit image) or from 0 to 4095 (a 12-bit image) for each color channel. For any size and bit depth of an image, we can compute the global information (Algorithm~\ref{Alg1}) provided by the occupied intensity bin $i$ and evaluate as a
change of a probability intensity histogram after removing a point from this bin.

For each parameter $\alpha$, the calculation of $\Gamma_{\alpha}^{(i)}$ helps to find values of the intensities with the identical occurrences and determine their distribution in (a structural part of) the image. Thus, in general, the recalculations to $\Gamma_{\alpha}^{(i)}$ can be considered as Look-Up Tables---intensities with the highest probabilities of occurrences in an image correspond to the highest (positive) values $\Gamma_{\alpha}^{(i)}$ and the brightest intensities in a $\Gamma_{\alpha}^{(i)}$-transformed image and vice versa. Sometimes, mainly in the case of local information, due to the transformation of the original values $\Gamma_{\alpha}^{(i)}$ into an 8-bit resolution, some levels $\Gamma_{\alpha}^{(i)}$ are merged into one intensity level of the transformed image.

\newpage

\begin{algorithm}[H]
\caption{Point information gain vector ($\mathbf{\Gamma}_\alpha$), point information gain entropy ($H_\alpha$), and point information gain entropy density ($\Xi_\alpha$) calculations for global (Whole image) information and typical histograms.}
\label{Alg1}
\LinesNumbered
\vspace{3pt}
\KwIn{$n$-bin histogram $\mathbf{h}$; $\alpha$, where $\alpha$ $\geq$ 0 $\land$ $\alpha$ $\neq$ 1}
\KwOut{$\mathbf{\Gamma}_\alpha$; $H_\alpha$; $\Xi_\alpha$}
\BlankLine
\BlankLine
	$\mathbf{p}=\mathbf{h}/$sum$(\mathbf{h})$; \qquad \emph{\% explain the frequency histogram $\mathbf{h}$ as a probability histogram $\mathbf{p}$}\\
	$\mathbf{\Gamma}_\alpha=$ zeros$(\mathbf{h})$; \qquad \emph{\% create a zero matrix $\mathbf{\Gamma}_\alpha$ of the size of the histogram $\mathbf{h}$}\
\BlankLine
\BlankLine
\For {$i = 1$ \KwTo $n$}{
	$\mathbf{h_2} = \mathbf{h}$; \qquad \emph{\% create a vector $\mathbf{h_2}$ identical to the histogram $\mathbf{h}$}\\
\BlankLine
\BlankLine
\If{$\mathbf{h_2}(i) \neq 0$}{
	$\mathbf{h_2}(i) = \mathbf{h_2}(i)-1$; \newline \emph{\% if the bin i of the histogram $\mathbf{h_2}$ is occupied, remove an element at the position i}\\
}
\BlankLine
\BlankLine
	$\mathbf{p_2} = \mathbf{h_2}/$sum$(\mathbf{h_2})$; \qquad \emph{\% calculate a probability histogram $\mathbf{p_2}$ without the examined element}\\
	$\mathbf{\Gamma}_\alpha^{(i)} = \frac{1}{1-\alpha}\log_2({\mbox{sum}(\mathbf{p_2}.^\land\alpha)}/{\mbox{sum}(\mathbf{p}.^\land\alpha))}$; \newline \emph{\% calculate $\Gamma_{\alpha}^{(i)}$ as a difference of two R\'{e}nyi entropies -- with and without the examined element, respectively Equation~\eqref{eg:1})}\\
}
\BlankLine
\BlankLine
$H_\alpha = \mbox{sum}(\mathbf{h}.^{*}\mathbf{\Gamma}_\alpha)$; \newline \emph{\% calculate $H_\alpha$ as a sum of the element-by-element multiplication of $\mathbf{h}$ and $\mathbf{\Gamma}_\alpha (Equation~\eqref{eg:6})$}\\
$\Xi_\alpha = \mbox{sum(}\mathbf{\Gamma}_\alpha)$; \qquad \emph{\% calculate $\Xi_\alpha$ as a sum of all unique values in  $\mathbf{\Gamma}_\alpha$ (Equation~\eqref{eg:7})}\;
\end{algorithm}

\begin{figure}[H]
\begin{center}
\includegraphics[width=.9\textwidth]{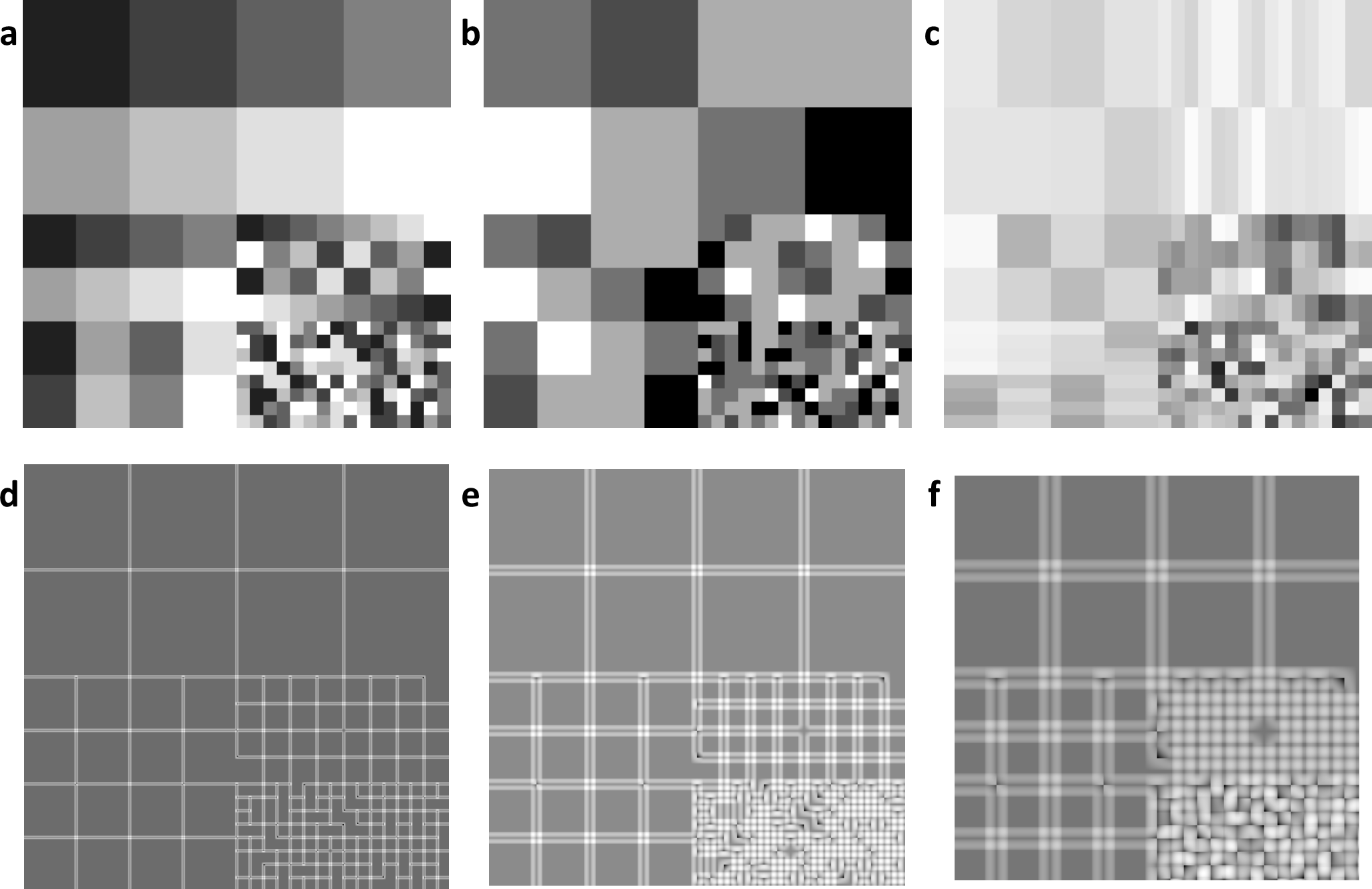}
\caption[]{$\Gamma_{0.99}^{(i)}$-transformations of the texmos2.s512 image \cite{USC}. Original image (\textbf{a}) and information images calculated from the whole image (\textbf{b}), a cross around each pixel (\textbf{c}), and squares of the side of 5, 15, and 29 px, respectively, with the centered examined pixel (\textbf{d--f}).}
\label{Fig3}
\end{center}
\end{figure}

\begin{figure}[H]
\begin{center}
\includegraphics[width=\textwidth]{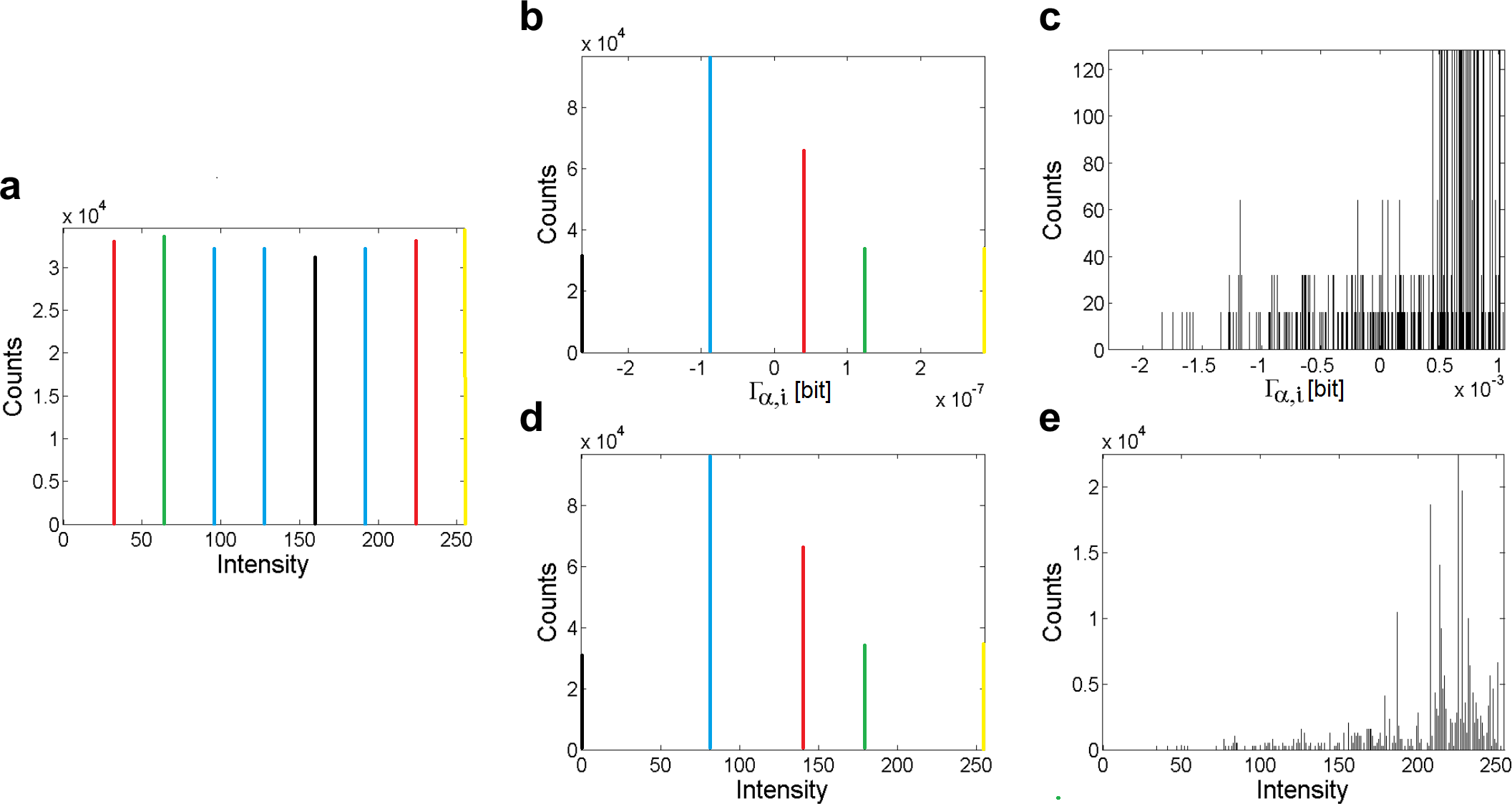}
\caption[] {Histograms of $\Gamma_{0.99}^{(i)}$-transformations of the texmos2.s512 image \cite{USC}. Original image (\textbf{a}), original values $\Gamma_{0.99}^{(i)}$ calculated from the whole image (\textbf{b}), original values $\Gamma_{0.99}^{(i)}$ calculated from a cross whose shanks intersect in the examined pixel (\textbf{c}), $\Gamma_{0.99}^{(i)}$-transformed images calculated from the whole image (\textbf{d}), and $\Gamma_{0.99}^{(i)}$-transformed images calculated from a cross around each pixel (\textbf{e}). Colors in the original and globally (whole image) transformed histograms correspond to the intensity levels with the identical frequencies of occurrences in the original image.}
\label{Fig4}
\end{center}
\end{figure}

\begin{figure}[H]
\begin{center}
\includegraphics[width=.9\textwidth]{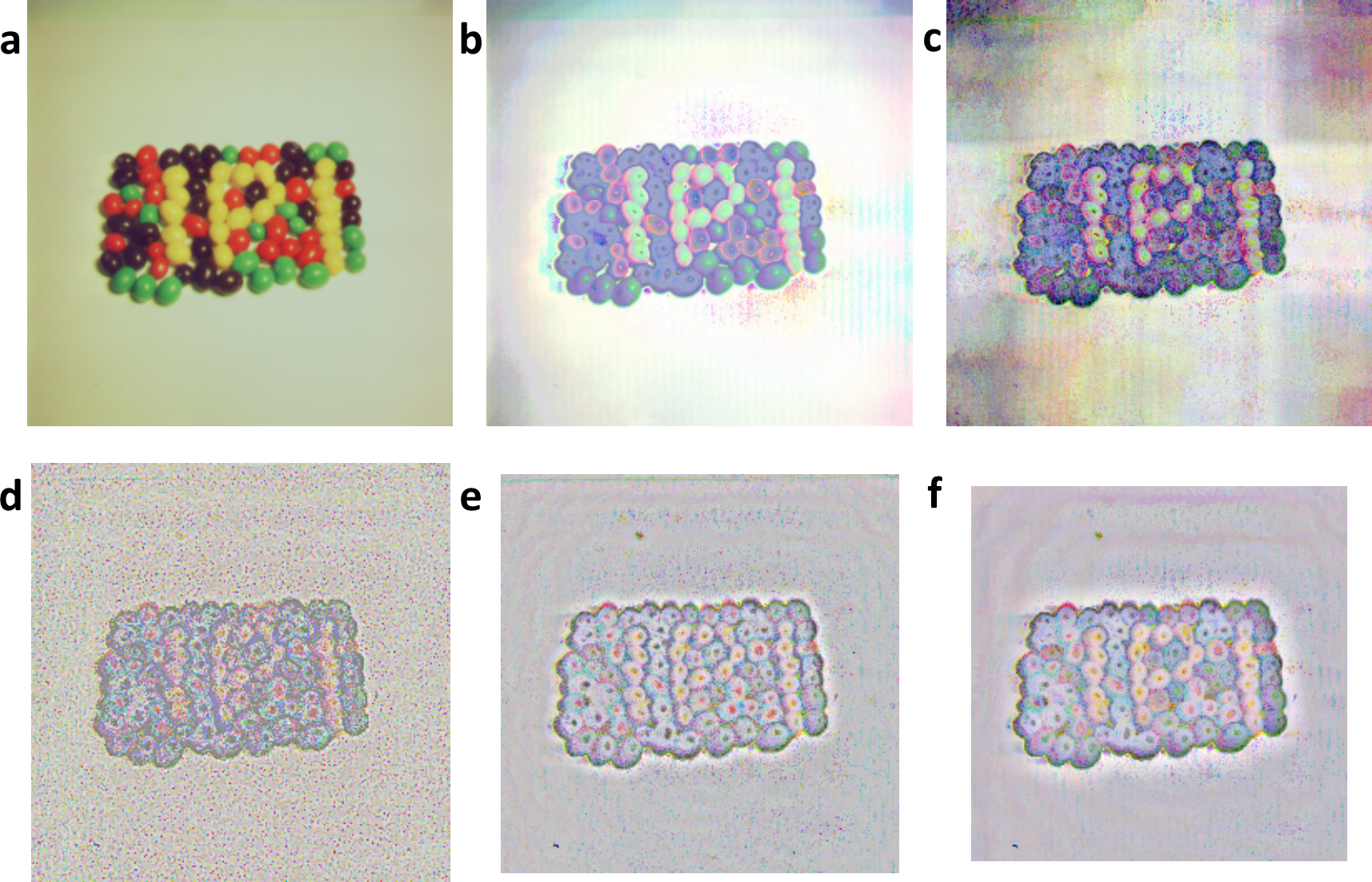}
\caption[]{$\Gamma_{0.99}^{(i)}$-transformations of the 4.1.07 image \cite{USC}. Original image (\textbf{a}) and information images calculated from the whole image (\textbf{b}), a cross around each pixel (\textbf{c}), and circles of the diameter of 5, 17, and 30 px, respectively, with the centered examined pixel (\textbf{d--f}).}
\label{Fig5}
\end{center}
\end{figure}

\vspace{6pt}
Everything is best visualized in Figures~\ref{Fig3}--\ref{Fig4}, which show the $\Gamma_{\alpha}^{(i)}$-transformations of the texmos2.s512  image. The intention was probably to create an image with a uniform distribution of intensities. Provided the uniform intensity distribution, the output of the global $\Gamma_{\alpha}^{(i)}$-calculation would be only one value $\Gamma_{\alpha}^{(i)}$, i.e., Figure~\ref{Fig3}b would be unicolor. However, eight original intensities (Figure~\ref{Fig3}a) resulted in five values $\Gamma_{0.99}^{(i)}$ (i.e., local parts) (Figure~\ref{Fig4}b,d). The detailed image analysis showed that the number of occurrences is only identical for intensities {32-224 and 96-128-192}, i.e.,
there are five unique values of frequencies of intensity occurrences (Figure~\ref{Fig4}a). For a change, in the 4.1.07 image, the global $\Gamma_{0.99}^{(i)}$-recalculation emphasizes the unevenness of the background and shadows around a group of the jelly beans (Figure~\ref{Fig5}b). In conformity with the statement in the next-to-last paragraph in Section~\ref{subsec:PIG}, this principle also enables highlighting of rare points in images with rich spectrum of intensities,
mainly at low $\alpha$-values. The calculations using higher values $\alpha$ do not point highlight rare points so intensively  and the resulting image is more smooth.

\subsection{Local Point Information Gain}
Since multidimensional datasets, as e.g., images, consist of special structures given by the pixel lattice, it can be also beneficial to calculate not only global information gain, but also local information gain in some defined surroundings (Algorithm~\ref{Alg2}). The local information is defined after removing an element from the bin $i$ where the element lies in the center of the surroundings, which creates the intensity histogram. The choice of the local surroundings around pixels is specific for each image. However, we do not have any systematic method for comparison of suitability of different surroundings around the pixels. The suitability of the chosen surroundings depends obviously on the process by which the observed pattern or other distribution was generated. According to our knowledge, the choice of the appropriate surroundings on the basis of known image generation was studied only for cellular automata~\cite{Shalizi2001, Shalizi_and_Shalizi2005, Shalizi_et_al_2002}. This makes the study of the local information very interesting because it outlines another method for recognition of the processes of self-organization/pattern formation~\cite{Crutchfield2012}. In this article, we confine ourselves to the usage of the local information for better understanding of both the limitation of the method of the $\Gamma_{\alpha}^{(i)}$-calculation and the local information itself. The cross, square, and circular surroundings around each pixel are demonstrated on three different standard images---texmos2.s512 (monochrome, computer-generated, unifractal), 4.1.07 (RGB, photograph, unifractal)~\cite{USC}, and wd950112 (monochrome version, computer-generated, multifractal)~\cite{pin}.

The cross from the intensity values, whose shanks meet in the examined point of the original image~\cite{Ur08}, was chosen as the first local surroundings. In contrast to the global recalculation, such a transformation of the texmos2.s512 image produces a substantially much richer intensity $\Gamma_{\alpha}^{(i)}$-image. One can see that relatively simple global information consists of more complex local information (Figure~\ref{Fig4}a,c,e).

However, the cross-local type of the image transformation is the least suitable approach for the analysis of the photograph of the jelly beans (Figure~\ref{Fig5}c). In this case, a circular local element is recommended to be used instead. As seen in Figure~\ref{Fig5}d--f,
the increase of the diameter up to the size of the jelly beans reduces the background gradually. The next increase enables grouping the jelly beans into higher-order assemblies. A similar grouping is observable for the smallest squares in the transformed texmos2.s512 using the 29 px square surroundings (Figure~\ref{Fig3}f). In contrast, lower values of square surroundings (Figure~\ref{Fig3}d) highlight only the border intensities.

\vspace{12pt}
\begin{algorithm}[H]
\label{Alg2}
\IncMargin{1em}
\LinesNumbered
 \caption{{Point information gain matrix} ($\mathbf{\Gamma}_\alpha$), point information gain entropy ($H_\alpha$), and point information gain entropy density ($\Xi_\alpha$) calculations for local kinds of information. Parameters $a$ and $b$ are semiaxes of the ellipse surroundings and a half-width of the rectangle surroundings, respectively, $a$ = 0 and $b$ = 0 for the cross surroundings.} \vspace{3pt}
\KwIn{{2D discrete data $\mathbf{I}_{m\times n}$}; $\alpha$, where $\alpha$ $\geq$ 0 $\land$ $\alpha$ $\neq$ 1; \mbox{parameters of surroundings} $a, b$}
\KwOut{$\mathbf{\Gamma}_{\alpha}^{(i)}$; $H_\alpha$; $\Xi_\alpha$}
\BlankLine
\BlankLine

$\mathbf{\Gamma}_\alpha = $ \mbox{zeros}$(\mathbf{I})$; \qquad \emph{\% create a zero matrix $\mathbf{\Gamma}_\alpha$ of the size of the $\mathbf{I}_{m\times n}$ matrix}\\
$\mathbf{hashMap} =$ \mbox{containers.Map}; \qquad \emph{\% declare an empty hash-map (the key-value array)}\\
\BlankLine

	\For {$i = (a+1)$ \KwTo $(m-a-1)$}{
		\For {$j = (b+1)$ \KwTo $(n-b-1)$}{
			$\mathbf{h} =$ \mbox{getHist}$(\mathbf{I}(i,j))$; \newline \emph{\% create a histogram $\mathbf{h}$ from the elements around the pixel (i,j) of $\mathbf{I}_{m\times n}$}\\
			$\mathbf{p}=\mathbf{h}/$sum$(\mathbf{h})$; \qquad \emph{\% explain the histogram $\mathbf{h}$ as a probability histogram $\mathbf{p}$}\\
			$\mathbf{h}(\mathbf{I}(i,j)) = \mathbf{h}(\mathbf{I}(i,j))-1$; \qquad \emph{\% remove the examined point (i,j) from the histogram $\mathbf{h}$}\\		
			$\mathbf{p_{2}}=\mathbf{h}/$sum$(\mathbf{h})$; \qquad \emph{\% explain the histogram $\mathbf{h}$ as a probability histogram $\mathbf{p_{2}}$}\\
			$\mathbf{\Gamma}_\alpha^{(i,j)}= \frac{1}{1-\alpha}\log_2({\mbox{sum}(\mathbf{p_2}.^\land\alpha)}/{\mbox{sum}(\mathbf{p}.^\land\alpha))}$; \newline \emph{\% calculate $\Gamma_{\alpha}^{(i,j)}$ as a difference of two R\'{e}nyi entropies -- with and without the examined element $(i,j)$ (Equation~\eqref{eg:1})}\\
\BlankLine
\BlankLine

$v = \mathbf{I}(i,j)$; \qquad \emph{\% read a value of the element (intensity) at the position $\mathbf{I}(i,j)$}\\
$\textit{checkSum} = \mbox{calcCheckSum}(\mathbf{h},v)$; \newline \emph{\% calculate \textit{checkSum} using a hash-function effective enough (MD4, MD5, SHA1,...)}\\
\BlankLine
\BlankLine

\If{$\mathrm{not}$ $\mathbf{hashMap}$$\mathrm{.isKey}$(\textit{checkSum})}{
	$\mathbf{hashMap}(\textit{checkSum}) = \mathbf{\Gamma}_\alpha(i,j)$; \newline \emph{\% if the hash-map does not contain the key, insert a new element with the key checkSum, where the inserted value is the $\Gamma_{\alpha}$ for the element $\mathbf{I}(i,j)$}\\
}
	}
	}
\BlankLine
\BlankLine

$H_\alpha = \mbox{sum}(\mbox{sum}(\mathbf{\Gamma}_\alpha))$; \qquad \emph{\% calculate $H_\alpha$ as a sum of all elements in the matrix $\mathbf{\Gamma}_\alpha$ (Equation~\eqref{eg:6})}\\
$\Xi_\alpha = \mbox{sum}(\mbox{values}(\mathbf{hashMap}))$; \qquad \emph{\% calculate $\Xi_\alpha$ as a sum of all elements in the matrix $\mathbf{hashMap}$ (Equation~\eqref{eg:7})}\\

\BlankLine
\BlankLine
\end{algorithm}

\subsection{Point Information Gain Entropy and Point Information Gain Entropy Density}
\begin{figure}[H]
\begin{center}
\includegraphics[width=0.9\textwidth]{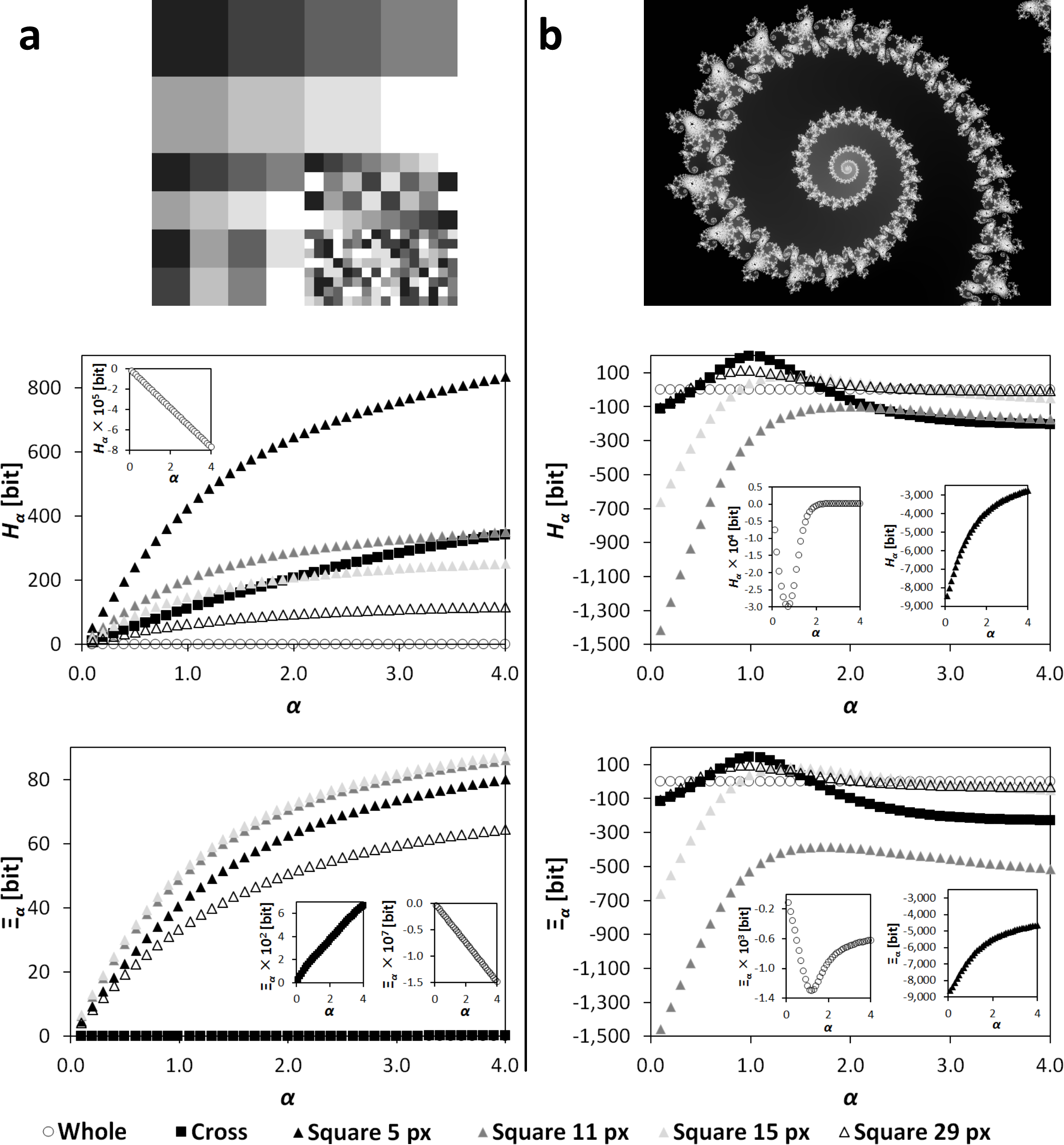}
\caption[]{Spectra $H_\alpha$ and $\Xi_\alpha$ for global information and different local surroundings of a unifractal (texmos2.s512 \cite{USC}, column (\textbf{a})) and multifracal (wd950112 \cite{pin}, column (\textbf{b})) image at \linebreak $\alpha$ = $\{0.1, 0.2,..., 0.9, 0.99, 1.1, 1.2,..., 4.0\}$.}
\label{Fig6}
\end{center}
\end{figure}

From the point of view of thermodynamics, the $H_\alpha$ and $\Xi_\alpha$ can be considered as additive, homological state variables whose knowledge can be helpful in analysis of multidimensional (image) data as well~\cite{zhyrova}. Despite the relative familiarity of their formulas (Section~\ref{subsec:PIE_and_PIED}), the $H_\alpha$ can be defined as a sum of all information contributions to the data distribution, either the global or partial one, i.e., all $\Gamma_{\alpha}^{(i)}$, whereas the $\Xi_\alpha$ is a sum of all information microstates of the distribution. Even in case of the local information, each two (collision) histograms with the same proportional representation of frequencies of elements, which were obtained from distributions around two pixels at different positions and only differing in the positions of frequencies in the histogram, are considered to be unique microstates and produce unique values $\Gamma_{\alpha}^{(i)}$ (see Algorithm~\ref{Alg2}). Thus, in agreement with the predictions arising from Equations~\eqref{eg:6} and \eqref{eg:7}, the $\Xi_\alpha$-calculation does not suppress contributions of elements with low probabilities of occurrences (rare points) and is more robust and stable against changes in the local surroundings. This phenomenon manifests itself in the lower differences in dependencies $\Xi_\alpha$($\alpha$) for four square surroundings in comparison to the dependencies $H_\alpha$($\alpha$) in Figure~\ref{Fig6}. Nevertheless, it is worth noting that, during the calculation with the usage of the local geometrical surroundings, the surroundings touch the edges of the image at most and only an interior part of the image is processed. This fact---technical limitation---negatively influences values $H_\alpha$ and $\Xi_\alpha$ for square surroundings in Figure~\ref{Fig6} and also leads to the lower sizes of $\Gamma_{\alpha}^{(i)}$-transformed images (e.g., Figures~\ref{Fig3}d--f and~\ref{Fig5}d--f).

Plotting the $H_\alpha$ and $\Xi_\alpha$ vs.~$\alpha$ in Figure~\ref{Fig6} is not random. As mentioned for $\Gamma_{\alpha}^{(i)}$ calculations (Section~\ref{subsec:PIG}), multidimensional discrete (image) data is suitable to be characterized not only by one discrete value, either $H_\alpha$ or $\Xi_\alpha$, at a particular $\alpha$, but also by their $\alpha$-dependent spectra. The reason is not only to avoid digital rounding, but also to possibly to characterize the type and the origin of geometrical structures in the image (cf. Section~\ref{subsec:PIGmultidimensional}). Another application has been found in the statistical evaluation (clustering) of the time-lapse multidimensional datasets~\cite{zhyrova,rychtarikova}. This calculation method was originally developed for study of multifractal self-organizing biological images~\cite{stysA,stysB}; however, it enables description of any types of images. Since parts of an image are forms of complex structures, the best way to interpret the image is to use a combination of its global and local kinds of information. We demonstrate this fact on an example of a unifractal (almost non-fractal) Euclidian image and a computer-generated multifractal image (Figure~\ref{Fig6}). Whereas the Euclidian image gives monotone spectra $H_\alpha$/$\Xi_\alpha$($\alpha$) (for the global and cross-local kinds of information, even 
linear dependencies
at the particular discrete interval of values $\alpha$), the recalculation of the multifractal image shows extremes at values of $\alpha$ close to 1. Analogous dependences were also plotted for the image sets of the course of the self-organizing Belousov--Zhabotinsky reaction~\cite{zhyrova}.

\section{Materials and Methods}
\label{methods}
\subsection{Processing of Images and Typical Histograms}
\label{image_process}
The values of $\Gamma_{\alpha}^{(i)}$, $H_\alpha$, and $\Xi_\alpha$ for all typical histograms and images were computed using Equations~\eqref{eg:1}, \eqref{eg:6}, and~\eqref{eg:7}. Algorithms are described in Section \ref{sec:algorithms}. The software and scripts, as well as results of all calculations, are available via ftp (Appendix).

For the Cauchy, L\'{e}vy, and Gauss distributions, histograms of dependences of the number of elements on the $\Gamma_{\alpha}^{(i)}$ were calculated for $\alpha$ = $\{$0.1, 0.3, 0.5, 0.7, 0.99, 1.3, 1.5, 1.7, 2.0, 2.5, 3.0, 3.5, 4.0$\}$ using a Matlab$^\circledR$ script. 
The following probability density functions $f(x)$ were studied:

\vspace{6pt}
\noindent(a) L\'{e}vy distribution:
\begin{equation}
f(x) = \mbox{round}\bigg[10^c\frac{\exp({-\frac{1}{2x}})}{\sqrt{2\pi x^3}}\bigg], \quad x \in \langle 1, 256\rangle, \quad x \in \mathbb{N}, \quad c \in \{3,5,7\},
\end{equation}

\noindent(b) Cauchy distribution:
\begin{equation}
f(x) = \mbox{round}\bigg[10^c\frac{1}{\pi (1+x^2)}\bigg], \quad x \in \langle 0, 255\rangle, \quad x \in \mathbb{Z}, \quad c \in \{3.5,7\},
\end{equation}

\noindent(c) Gauss distribution:
\begin{equation}
f(x) = \mbox{round}\bigg[10^c\frac{\exp({-\frac{x^2}{2\sigma^2}})}{\sigma \sqrt{2\pi}}\bigg], \quad x \in \langle 0, 255\rangle, \quad x \in \mathbb{Z}, \quad c \in \{4, 300\} \land \sigma = 1,\, c \in \{3, 4\} \land \sigma = 10.
\end{equation}
In Figures~\ref{Fig1} and \ref{Fig2}, the Cauchy and L\'{e}vy distributions with $c$ = 7 and the Gauss distribution with parameters $c$ = 4 and $\sigma$ = 10 are depicted.

Multidimensional image analysis based on calculation of $\Gamma_{\alpha}^{(i)}$, $H_\alpha$, and $\Xi_\alpha$ was tested on 5 standard 8-{bpc} images (Table~\ref{Tab1}). Before the computations, original images wd950112.gif and
6ASCP011.gif obtained from~\cite{pin} were transformed into monochrome *.png formats in Matlab$^\circledR$ software. All images were processed using an Image Info Extractor Professional software (Institute of Complex System,  University of South Bohemia, Nov\'{e} Hrady,
 Czech Republic) for $\alpha$ = $\{$0.1, 0.2, ..., 0.9, 0.99, 1.1, 1.2, ..., 4.0$\}$. The global
information was extracted using (the italics refer to parameters which are set in the Image Info Extractor Professional software.) 
 $Whole$ $Image$ calculation. The vertical--horizontal cross, square (a side of 5, 11, 15, and 29 px, respectively), and circle (a radius of 2, 5, and 8 px, respectively) for local information were set as special cases of a $Cross$, $Rectangle$, and $Ellipse$ calculation at the rotation angle $Phi$ of 0$^\circ$. Into the Image Info Extractor Professional software, a side of the square and radius of the circle surroundings was input as $width/2$ and $height/2$ of 2, 5, and 14 px and $a$ and $b$ of 2, 5, and 8 px, respectively.
\begin{table}[H]
\caption{Specifications of images.}
\label{Tab1}
\centering
\begin{tabular}{rccccc}
\toprule
 \textbf{Image} & \textbf{Source} & \textbf{Colors} & \textbf{Resolution} & \textbf{Geometry} & \textbf{Origin}\\
\midrule
    texmos2.s512.png & \cite{USC} & mono & 512$\times$512 & unifractal & computer-based\\
    4.1.07.tiff & \cite{USC} & RGB & 256$\times$256 & unifractal & photograph\\
    wash-ir.tiff & \cite{USC} & RGB & 2250$\times$2250 & unifractal & computer-based\\
    wd950112.png & \cite{pin} & mono & 1024$\times$768 & multifractal & computer-based\\
    6ASCP011.png & \cite{nyu} & mono & 1600$\times$1200 & multifractal & computer-based\\
\bottomrule
\end{tabular}
\end{table}

\subsection{Calculation Algorithms}
\label{sec:algorithms}
The algorithms implemented into the Image Info Extractor Professional are described in Algorithms \ref{Alg1}--\ref{Alg2}. In the case of RGB images, the algorithms were applied to each color channel. The values $\Gamma_{\alpha}^{(i)}$ were visualized by a full rescaling into 8-bit resolution. Let us note that, for $\alpha$ = 1, the equations in lines 9 of both algorithms switch to the calculation of the Shannon entropy.

\section{Conclusions}
\label{}
In this article, we propose novel information quantities---a point information gain ($\Gamma_{\alpha}^{(i)}$), a point information gain entropy ($H_\alpha$), and a point information gain entropy density ($\Xi_\alpha$). We found a monotone dependency of the number of the elements of a given property in the set on $\Gamma_{\alpha}^{(i)}$. The variables $H_\alpha$ and $\Xi_\alpha$ can be used as quantities in multidimensional datasets for the definition of the information context. Examination of local information in the distribution shows a potential for in-depth insight into formation of observed structures and patterns. This option can be practically utilized in acquisition of differently resolved variables in the dataset. The method enables avoiding cases where the number of occurrences of a certain event is the same, but ,in distribution in time, space or along any other variable, differ. In principle, the variables $H_\alpha$ and $\Xi_\alpha$ are unique for each distribution but suffer from problems with digital precision of the computation. Therefore, we propose their $\alpha$-dependent spectra as proper characteristics of any discrete distribution, e.g., for clustering of multidimensional datasets.

\acknowledgments{This work was supported by the Ministry of Education, Youth and Sports of the Czech Republic---projects CENAKVA (No. CZ.1.05/2.1.00/01.0024), CENAKVA II (No. LO1205 under the NPU I program), and the
CENAKVA Centre Development (No. CZ.1.05/2.1.00/19.0380). Jan Korbel acknowledges the support from the Czech Science Foundation, Grant No. GA14-07983S.}

\authorcontributions{Renata Rycht\'{a}rikov\'{a} was the main author of the text and tested the algorithms; Jan~Korbel was responsible for the theoretical part of the article; Petr Mach\'{a}\v{c}ek and Petr C\'{i}sa\v{r} were the developers of the Image Info Extractor Professional software; Jan Urban was the first who derived the point information gain from Shannon entropy; Dalibor \v{S}tys was the group leader who derived the point information gain for the R\'{e}nyi entropy and prepared the first version of the manuscript. All authors have read and approved the final manuscript.}

\conflictofinterests{The authors declare no conflict of interest.}


\appendix
\section{}
All processed data are available
~at~\cite{suppl} (for more details, see Section~\ref{methods}):
\begin{enumerate}[leftmargin=*,labelsep=3mm]
\item Folder ``Figures'' contains subfolders with results of $\Gamma_{\alpha}^{(i)}$, $H_\alpha$, and $\Xi_\alpha$ calculations for ``RGB'' (4.1.07.tiff, wash-ir.tiff) and ``gray'' (texmos2.s512.png, wd950112.png, 6ASCP011.png) standard images calculated for 40 values $\alpha$. The results are separated into subfolders according to the type of extracted information.
\item Folder ``H$\_$Xi'' stores the PIE$\_$PIED.xlsx and PIE$\_$PIED2.xlsx files with dependencies of $H_\alpha$ and $\Xi_\alpha$ on $\alpha$ as exported from the PIE.mat files (in folder ``Figures''). Titles of the graphs, which are in agreement with the computed variables and extracted kinds of information, are written in the~sheets.
\item Folder ``Histograms'' stores the histograms of the occurrences of $\Gamma_{\alpha}^{(i)}$ values for the Cauchy (two types), L\'{e}vy (three types), and Gauss (four types) distributions. The parameters of the original distributions are saved in the equation.txt files. All histograms were recalculated using 13 values~$\alpha$.
\item Folder ``Software'' contains a 32- and 64-bit version of an Image Info Extractor Professional v. b9 software (ImageExtractor$\_$b9$\_$xxbit.zip; supported by OS Win7) and a pig$\_$histograms.m Matlab$^\circledR$ script for recalculation of the typical probability density functions. A script pie$\_$ec.m serves for the extraction of $H_\alpha$ and $\Xi_\alpha$ from the folders (outputs from the Image Info Extractor Professional) over $\alpha$. In the software and script, the variables $\Gamma_{\alpha}^{(i)}$, $H_\alpha$, and $\Xi_\alpha$ are called $PIG$, $PIE$, and $PIED$, respectively. Manuals for the software and scripts are also attached.
\end{enumerate}

\bibliographystyle{mdpi}

\begin{thebibliography}{999}

\providecommand{\natexlab}[1]{#1}

\bibitem[\v{S}tys \em{et~al.}(2010)\v{S}tys, Urban, Van\v{e}k, and
  C\'{i}sa\v{r}]{Ur08}
\v{S}tys, D.; Urban, J.; Van\v{e}k, J.; C\'{i}sa\v{r}, P.
\newblock Analysis of biological time-lapse microscopic experiment from the
  point of view of the information theory.
\newblock {\em Micron} {\bf 2011}, {\em 42},~3360--3365.

\bibitem[Urban \em{et~al.}(2009)Urban, Van\v{e}k, and \v{S}tys]{Ur09}
Urban, J.; Van\v{e}k, J.; \v{S}tys, D.
\newblock Preprocessing of microscopy images via Shannon's entropy.
\newblock  In Proceedings of Pattern Recognition and Information Processing,
  Minsk, Belarus,  19--21  May 2009; pp. 183--187.

\bibitem[Boer \em{et~al.}(2002)Boer, Kroese, Mannor, and Rubinstein]{cross}
Boer, P.T.D.; Kroese, D.P.; Mannor, S.; Rubinstein, R.
\newblock A tutorial on the cross-entropy method.
\newblock {\em Ann. Oper. Res.} {\bf 2005}, {\em 134}, 19--67.

\bibitem[Baez \em{et~al.}(2011)Baez, Fritz, and Leinster]{baez2011}
Baez, J.C.; Fritz, T.; Leinster, T.
\newblock A characterization of entropy in terms of information loss.
\newblock {\em Entropy} {\bf 2011}, {\em 13},~1945--1957.

\bibitem[Marcolli and Tedeschi(2015)]{marcolli2015}
Marcolli, M.; Tedeschi, N.
\newblock Entropy algebras and Birkhoff factorization.
\newblock {\em J. Geom. Phys.} {\bf 2015}, {\em 97},~243--265.

\bibitem[Tsallis(1988)]{Ts88}
Tsallis, C.
\newblock Possible generalization of Boltzmann--Gibbs statistics.
\newblock {\em J. Stat. Phys.} {\bf 1988}, {\em 52},~479--487.

\bibitem[Jizba and Arimitsu(2004)]{Ji04a}
Jizba, P.; Arimitsu, T.
\newblock The world according to R\'{e}nyi: Thermodynamics of multifractal
  systems.
\newblock {\em Ann. Phys.} {\bf 2004}, {\em 312},~17--59.

\bibitem[Jizba and Korbel(2014)]{Jizba14}
Jizba, P.; Korbel, J.
\newblock Multifractal diffusion entropy analysis: Optimal bin width of
  probability histograms.
\newblock {\em Physica A} {\bf 2014}, {\em 413},~438--458.

\bibitem[Hentschel and Procaccia(1985)]{He85}
Hentschel, H.G.E.; Procaccia, I.
\newblock The infinite number of generalized dimensions of fractals and strange
  attractors.
\newblock {\em Physcia D} {\bf 1983}, {\em 8},~435--444.

\bibitem[Campbel(1965)]{Ca65}
Campbel, L.L.
\newblock A coding theorem and R\'{e}nyi's entropy.
\newblock {\em Inf. Control} {\bf 1965}, {\em 8},~423--429.

\bibitem[\v{S}tys \em{et~al.}(2012)\v{S}tys, Jizba, Pap\'{a}\v{c}ek,
  N\'{a}hlik, and C\'{i}sa\v{r}]{St12}
\v{S}tys, D.; Jizba, P.; Pap\'{a}\v{c}ek, S.; N\'{a}hlik, T.; C\'{i}sa\v{r}, P.
\newblock On measurement of internal variables of complex self-organized
  systems and their relation to multifractal spectra.
\newblock  In Proceedings of the 6th IFIP TC 6 International Workshop (WSOS 2012), Delft, The Netherlands, 15--16 March 2012; pp. 36--47.

\bibitem[R\'{e}nyi(1961)]{Re61}
R\'{e}nyi, A.
\newblock On measures of entropy and information.
In Proceedings of the Fourth Berkeley Symposium on Mathematical Statistics and Probability, Berkeley, CA, USA, 20 June--30 July 1960; pp. 547--561.

\bibitem[Kullback and Leibler(1951)]{Ku57}
Kullback, S.; Leibler, R.A.
\newblock On information and sufficiency.
\newblock {\em Ann. Math. Stat.} {\bf 1951}, {\em 22},~79--86.

\bibitem[Csisz\'{a}r(1975)]{Cs75}
Csisz\'{a}r, I.
\newblock $I$-divergence geometry of probability distributions and minimization
  problems.
\newblock {\em Ann. Prob.} {\bf 1975}, {\em 3},~146--158.

\bibitem[Harremoes(2006)]{Ha06}
Harremoes, P.
\newblock Interpretations of R\'{e}nyi entropies and divergences.
\newblock {\em Physica A} {\bf 2006}, {\em 365},~57--62.

\bibitem[van Erven and Harremoes(2014)]{Er07}
Van Erven, T.; Harremoes, P.
\newblock R\'{e}nyi divergence and Kullback--Leibler divergence.
\newblock {\em IEEE Trans. Inf. Theory} {\bf 2014}, {\em 60},~3797--3820.

\bibitem[van Erven and Harremo{\"e}s(2010)]{Er10}
Van Erven, T.; Harremo{\"e}s, P.
\newblock R{\'e}nyi divergence and majorization.
\newblock  In Proceedings of the 2010 IEEE International
  Symposium on Information Theory Proceedings (ISIT), Austin, TX, USA, 13--18 June 2010.

\bibitem[Jizba \em{et~al.}(2012)Jizba, Kleinert, and Shefaat]{Ji12}
Jizba, P.; Kleinert, H.; Shefaat, M.
\newblock R\'{e}nyi's information transfer between financial time series.
\newblock {\em Physica A} {\bf 2012}, {\em 391},~2971--2989.

\bibitem[Havrda and Charv\'{a}t(1967)]{Ha67}
Havrda, J.; Charv\'{a}t, F.
\newblock Quantification method of classification processes. Concept of
  structural $\alpha$-entropy.
\newblock {\em Kybernetika} {\bf 1967}, {\em 3},~30--35.

\bibitem[Grassberger and Procaccia(1983{\natexlab{a}})]{Gr83a}
Grassberger, P.; Procaccia, I.
\newblock Measuring the strangeness of strange attractors.
\newblock {\em Physica D} {\bf 1983}, {\em 9},~189--208.

\bibitem[Grassberger and Procaccia(1983{\natexlab{b}})]{Gr83b}
Grassberger, P.; Procaccia, I.
\newblock Characterization of strange attractors.
\newblock {\em Phys. Rev. Lett.} {\bf 1983}, {\em 50},~346.

\bibitem[Costa(1985)]{Costa1985}
Costa, M.
\newblock A new entropy power inequality.
\newblock {\em IEEE Trans. Inf. Theory} {\bf 1985}, {\em 31},~751--760.

\bibitem[Jizba \em{et~al.}(2015)Jizba, Dunningham, and Joo]{Jizbaetal2014}
Jizba, P.; Dunningham, J.A.; Joo, J.
\newblock Role of information theoretic uncertainty relations in quantum
  theory.
\newblock {\em Ann. Phys.} {\bf 2015}, {\em 355},~87--114.

\bibitem[Shannon(1948)]{Sh48}
Shannon, C.E.
\newblock A mathematical theory of communication.
\newblock {\em Bell Syst. Tech. J.} {\bf 1948}, {\em 27},~379--423, 623--656.

\bibitem[Jizba and Arimitsu(2004)]{Ji04b}
Jizba, P.; Arimitsu, T.
\newblock On observability of R\'{e}nyi's entropy.
\newblock {\em Phys. Rev. E} {\bf 2004}, {\em 69},~026128.

\bibitem[USC()]{USC}
The USC-SIPI Image Database. Available online: \url{http://sipi.usc.edu/database/database.php?volume=textures \&image=61\#top} (accessed on 17 October 2016).


\bibitem[Shalizi and Crutchfield(2011)]{Shalizi2001}
Shalizi, C.R.; Crutchfield, J.P.
\newblock Computational mechanics: Pattern and prediction, structure and
  simplicity.
\newblock {\em J. Stat. Phys.} {\bf 2001}, {\em 104},~817--879.

\bibitem[Shalizi and Shalizi(2003)]{Shalizi_and_Shalizi2005}
Shalizi, C.R.; Shalizi, K.L.
\newblock Quantifying self-organization in cyclic cellular automata.
\newblock  In {\em Noise in Complex Systems and Stochastic Dynamics}; Society of Photo Optical: Bellingham, WA, USA,  2003.

\bibitem[Shalizi \em{et~al.}(2004)Shalizi, Shalizi, and
  Haslinger]{Shalizi_et_al_2002}
Shalizi, C.R.; Shalizi, K.L.; Haslinger, R.
\newblock Quantifying self-organization with optimal predictors.
\newblock {\em Phys.~Rev.~Lett.} {\bf 2004}, {\em 93},~118701.



\bibitem[Crutchfield(2012)]{Crutchfield2012}
Crutchfield, J.P.
\newblock Between order and chaos.
\newblock {\em Nat. Phys.} {\bf 2012}, {\em 8},~17--24.


\bibitem[pin()]{pin}
\mbox{Explore Fractals Beautiful, Colorful Fractals, and More! Available online: https://www.pinterest.com/pin/}\\ 254031235202385248/ (accessed on 17 October 2016).

\bibitem[Zhyrova \em{et~al.}(2014)Zhyrova, \v{S}tys, and
  C\'{i}sa\v{r}]{zhyrova}
Zhyrova, A.; \v{S}tys, D.; C\'{i}sa\v{r}, P.
\newblock Macroscopic description of complex self-organizing system:
  Belousov--Zhabotinsky reaction. In {\em ISCS 2013: Interdisciplinary Symposium on Complex Systems}; Sanayei,~A., Zelinka, N., Rössler, O.E., Eds.; Springer: Berlin/Heidelberg,
  Germany,  2014; pp. 109--115.

\bibitem[Rychtarikova(2016)]{rychtarikova}
Rychtarikova, R.
\newblock Clustering of multi-image sets using R\'{e}nyi information entropy.
In {\em Bioinformatics and Biomedical Engineering}; Ortuño, F., Rojas, I., Eds.; Springer: Cham, Switzerland,  2016; pp. 517--526.

\bibitem[\v{S}tys \em{et~al.}(2011{\natexlab{a}})\v{S}tys, Urban, Van\v{e}k,
  and C\'{i}sa\v{r}]{stysA}
\v{S}tys, D.; Urban, J.; Van\v{e}k, J.; C\'{i}sa\v{r}, P.
\newblock Analysis of biological time-lapse microscopic experiment from the
  point of view of system theory.
\newblock {\em Micron} {\bf 2011}, {\em 42},~360--365.

\bibitem[\v{S}tys \em{et~al.}(2011{\natexlab{b}})\v{S}tys, Van\v{e}k,
  N\'{a}hl\'{i}k, Urban, and C\'{i}sa\v{r}]{stysB}
\v{S}tys, D.; Van\v{e}k, J.; N\'{a}hl\'{i}k, T.; Urban, J.; C\'{i}sa\v{r}, P.
\newblock The cell monolayer trajectory from the system state point of view.
\newblock {\em Mol. BioSyst.} {\bf 2011}, {\em 42},~2824--2833.

\bibitem[nyu()]{nyu}
Available online: http://cims.nyu.edu/~kiryl/Photos/Fractals1/ascp011et.html (accessed on 17 October 2016).

\bibitem[sup()]{suppl}
Point information gain supplementary data. Available online: ftp://160.217.215.251/pig
(user: anonymous; password: anonymous).

\end{thebibliography}

\renewcommand\bibname{References}

\end{document}